\documentstyle[preprint,epsfig,aps]{revtex}
\tightenlines
\begin{document}
\preprint{\vbox{\hbox{IFUSP/P-1475} 
                \hbox{UH-511-986-01}}}

\title{Probing Anomalous Quartic Couplings in $e\gamma$ and $\gamma\gamma$
Colliders}

\author{O.\ J.\ P.\ \'Eboli$^1$\thanks{Email: eboli@fma.if.usp.br;
    $^\dagger$ Email: mizuka@phys.hawaii.edu} and J.\ K.\
    Mizukoshi$^{2,\dagger}$}

\address{$^1$ Instituto de F\'{\i}sica, Universidade de S\~ao Paulo \\
              C.P.\ 66318, 05315--970, S\~ao Paulo, Brazil.}

\address{$^2$ Department of Physics $\&$ Astronomy, University of Hawaii\\
Honolulu, HI 96822, USA.}


\maketitle
\vskip -0.5cm
\begin{abstract} 
  
  We analyze the potential of the $e^+e^-$ Linear Colliders, operating in the
  $e\gamma$ and $\gamma\gamma$ modes, to probe anomalous quartic vector--boson
  interactions through the multiple production of $W$'s and $Z$'s. We examine
  all $SU(2)_L \otimes U(1)_Y$ chiral operators of order $p^4$ that lead to
  new four--gauge--boson interactions but do not alter trilinear vertices. We
  show that the $e\gamma$ and $\gamma\gamma$ modes are able not only to
  establish the existence of a strongly interacting symmetry breaking sector
  but also to probe for anomalous quartic couplings of the order of $10^{-2}$
  at 90\% CL. Moreover, the information gathered in the $e \gamma$ mode can be
  used to reduced the ambiguities of the $e^+e^-$ mode.

\end{abstract}


\section{Introduction}

The $SU_L(2) \times U_Y(1)$ local gauge symmetry of the Standard Model (SM)
determines completely the triple and quartic vector--boson interactions.
Therefore, the direct study of these couplings can further confirm the SM or
give some hint on the existence of new phenomena at a higher scale.  Moreover,
it is important to independently measure the trilinear and quartic gauge boson
couplings because there are extensions and limits of the SM \cite{hil:bes}
that leave the trilinear couplings unchanged but do modify the quartic
vertices.  Presently, the triple gauge--boson couplings are being probed at
the Tevatron \cite{teva} and LEP \cite{lep} through the production of vector
boson pairs, however, we have only started to study directly the quartic
gauge--boson couplings \cite{exp:LEP,lep:ichep}.  Due to the limited available
center--of--mass energy, the first quartic couplings to be studied contain two
photons, and just at the CERN Large Hadron Collider (LHC) and the next
generation of $e^+e^-$ Linear Colliders (LC) we will be able to probe $VVVV$
($V=W$ or $Z$) vertices.

If the $SU(2)_L \otimes U(1)_Y$ symmetry of the model is to be linearly
realized, studies of the triple gauge--boson couplings will be able to furnish
information on the gauge--boson four--point functions provided that dimension
8 and higher anomalous operators are suppressed. This is the case when the
breaking of the $SU(2)_L \otimes U(1)_Y$ symmetry takes place via the Higgs
mechanism with a relatively light elementary Higgs boson. If, on the other
hand, no fundamental light Higgs particle is present in the theory, one is led
to consider the most general effective Lagrangian which employs a nonlinear
representation of the broken $SU(2)_L \otimes U(1)_Y$ gauge symmetry
\cite{Appelquist}.  In this case the SM relation between the structure of the
three-- and four--point functions of the gauge bosons does not hold already at
$p^4$ order, leaving open the question of the structure of the quartic
vector--boson interactions.

LC provides a unique possibility to study $e\gamma$ and $\gamma\gamma$
collisions since high energy photons can be produced by laser backscattering
\cite{telnov}. These new modes of operation of the LC allow us to probe the
$W^+W^-W^+W^-$ and $W^+W^-ZZ$ couplings through the reactions 
\begin{eqnarray}
e^- \gamma &\to& W^- W^+ W^- \nu_e  \;\; ,
\label{wwwn} \\
e^- \gamma &\to& W^- Z Z \nu_e  \;\; ,
\label{wzzn} \\
\gamma\gamma &\to&  W^- W^+ W^- W^+  \;\; ,
\label{wwww} \\
\gamma\gamma &\to& W^- W^+ Z Z  \;\; ,
\label{wwzz}
\end{eqnarray}
which take place via weak boson fusion at high energies \cite{hwz}. In
this work, we access the reach of the LC operating in the $e\gamma$
and $\gamma\gamma$ modes to study the symmetry breaking sector via the
measurement of quartic gauge couplings.  We work in the framework of
chiral Lagrangians, and we analyze all $p^4$ operators that lead to
genuine quartic gauge interactions, {\em i.e.} these operators do not
give rise to triple gauge--boson vertices, and consequently are not
bounded by the study of the production of gauge--boson pairs. We also
take into account realistic cuts, detection efficiencies and potential
backgrounds.

At present the only information on quartic couplings $WWWW$ and $WWZZ$ is
obtained indirectly as they modify the gauge--boson two--point functions at
one loop \cite{adri}.  The precise electroweak measurements both at low energy
and at the $Z$ pole, constrain the quartic anomalous couplings to be smaller
than $10^{-3}$--$10^{-1}$ depending on the coupling.  In the future, quartic
interactions can be studied at the LHC through the reaction $pp \rightarrow V
V X$ \cite{pp} while the following processes can give information on these
couplings at the LC: $e^+ e^- \rightarrow VVV$ \cite{bhp}, $e^+ e^- 
\rightarrow F F VV$ \cite{bhk}, $e\gamma \rightarrow V V F$ \cite{our:eg}, 
$\gamma\gamma \rightarrow VV$ \cite{bela2}, and $\gamma \gamma \rightarrow VVV
$ \cite{our:vvv}, where $V=$ $Z$, $W^\pm$ or $\gamma$ and $F=$ $e$ or $\nu_e$.

In this work, we show that the LC operating in the $e\gamma$ or $\gamma\gamma$
modes not only can establish the existence of a strongly interacting symmetry
breaking sector but also can lead to bounds on genuine quartic interactions of
the order of $10^{-2}$ which turn out to be of the same order of magnitude of
the attainable limits at the LHC and $e^+e^-$ LC. Furthermore, the information
gathered in the $e \gamma$ mode can be used to reduced the ambiguities of the
$e^+e^-$ mode, which exhibits two allowed regions, leading to a better
determination of the quartic couplings.

The outline of this work is as follows. In Sec.\ \ref{theo} we present
the chiral lagrangian formalism that we employed. We describe the
calculational tools used in Sec.\ \ref{tools}, while the main features
of the signal and backgrounds are discussed in Sec.\ \ref{prop}. We
present our results in Sec.\ \ref{resu} which also contains our
conclusions.
 
\section{Theoretical Framework}
\label{theo}

If the electroweak symmetry breaking is due to a heavy (strongly interacting)
Higgs boson, which can be effectively removed from the physical low--energy
spectrum, or to no fundamental Higgs scalar at all, one is led to consider the
most general effective Lagrangian which employs a nonlinear representation of
the broken $SU(2)_L \otimes U(1)_Y$ gauge symmetry \cite{Appelquist}.  The
resulting chiral Lagrangian is a non--renormalizable non--linear $\sigma$
model coupled in a gauge--invariant way to the Yang--Mills theory.  This model
independent approach incorporates by construction the low-energy theorems
\cite{cgg}, that predict the leading behavior of Goldstone boson amplitudes
irrespective of the details of the symmetry breaking mechanism.
Notwithstanding, unitarity implies that this low--energy effective
theory should be valid up to some energy scale smaller than $4\pi v
\simeq 3$ TeV \cite{lee}, where new physics would come into play.

To specify the effective Lagrangian one must first fix the symmetry
breaking pattern. We consider that the system presents a global
$SU(2)_L \otimes SU(2)_R$ symmetry that is broken to $SU(2)_C$. With
this choice, the building block of the chiral Lagrangian, in the
notation of Ref.\ \cite{Appelquist}, is the dimensionless unimodular
matrix field $\Sigma(x)$, which transforms under $SU(2)_L \otimes
SU(2)_R$ as $(2,2)$:
\begin{equation}
\Sigma(x) ~=~ \exp\left(i \frac{\varphi^a(x) \tau^a}{v}\right) \; .
\end{equation}
The $\varphi^a$ fields are the would-be Goldstone fields and $\tau^a$
($a=1$, $2$, $3$) are the Pauli matrices.  The $SU(2)_L
\otimes U(1)_Y$ covariant derivative of $\Sigma$ is defined as
\begin{equation}
D_\mu \Sigma ~\equiv~ \partial_\mu \Sigma 
+ i g \frac{\tau^a}{2} W^a_\mu \Sigma -
i g^\prime \Sigma \frac{\tau^3}{2} B_\mu \; .
\end{equation}

The lowest-order terms in the derivative expansion of the effective
Lagrangian are
\begin{equation}
{\cal L}^{(2)} = \frac{v^2}{4} \hbox{Tr} \left [ \left ( D_\mu \Sigma \right )
^\dagger \left ( D^\mu \Sigma \right ) \right ]
+ \beta_1 g'^2\frac{v^2}{4} \left ( \hbox{Tr}
\left [ T V_\mu \right ] \right )^2
\; .
\label{lagran2}
\end{equation}
where we have introduced the auxiliary quantities $T \equiv\Sigma \tau^3
\Sigma^\dagger$ and $V_\mu \equiv \left ( D_\mu \Sigma \right ) \Sigma^\dagger
$ which are $SU(2)_L$--covariant and $U(1)_Y$--invariant. Notice that $T$ is
not invariant under $SU(2)_C$ custodial due to the presence of $\tau^3$.

The first term in Eq.\ (\ref{lagran2}) is responsible for giving mass to the
$W^\pm$ and $Z$ gauge bosons for $ v = ( \sqrt{2} G_F )^{-1} $.  The second
term violates the custodial $SU(2)_C$ symmetry and contributes to $\Delta\rho$
at tree level, being strongly constrained by the low--energy data. This term
can be understood as the low-energy remnant of a high--energy custodial
symmetry breaking physics, which has been integrated out above a certain scale
$\Lambda$.  Moreover, at the one--loop level, this term is also required in
order to cancel the divergences in $\Delta\rho$, arising from diagrams
containing a hypercharge boson in the loop.  This subtraction renders
$\Delta\rho$ finite, although dependent on the renormalization scale
\cite{Appelquist}.

At the next order in the derivative expansion, $D=4$, several operators can be
written down \cite{Appelquist}. We shall restrict ourselves to those
containing genuine quartic vector-boson interactions, which are
\begin{eqnarray}
{\cal L}^{(4)}_4 &=& \alpha_4\left[{\rm{Tr}}
\left(V_{\mu}V_{\nu}\right)\right]^2
\label{eff:4}
\;, \\
{\cal L}^{(4)}_5 &=& \alpha_5\left[{\rm{Tr}}
\left(V_{\mu}V^{\mu}\right)\right]^2
\label{eff:5}
\;, \\
{\cal L}^{(4)}_6 &=& \alpha_6 \; {\rm{Tr}}\left(V_{\mu}V_{\nu}\right)
{\rm{Tr}}
\left(TV^{\mu}\right){\rm{Tr}}\left(TV^{\nu}\right) 
\label{eff:6}
\;, \\
{\cal L}^{(4)}_7 &=& \alpha_7\;{\rm{Tr}}\left(V_{\mu}V^{\mu}\right)
\left[{\rm{Tr}}\left(TV^{\nu}\right)\right]^2
\label{eff:7}
\;, \\
{\cal L}^{(4)}_{10} &=& \frac{1}{2}
\alpha_{10}\left[{\rm{Tr}}\left(TV_{\mu}\right)
\;{\rm{Tr}}\left(TV_{\nu}\right)\right]^2
\; .
\label{eff:10}
\end{eqnarray}
In an arbitrary gauge, these Lagrangian densities lead to quartic vertices
involving gauge bosons and/or Goldstone bosons. In the unitary gauge, these
effective operators give rise to anomalous $ZZZZ$ (all operators), $W^+W^-ZZ$
(all operators except ${\cal L}^{(4)}_{10}$), and $W^+ W^- W^+ W^-$ (${\cal
L}^{(4)}_4$ and ${\cal L}^{(4)}_5$) interactions.  Moreover, the interaction
Lagrangians ${\cal L}^{(4)}_6$, ${\cal L}^{(4)}_7$, and ${\cal L}^{(4)}_{10}$
violate the $SU(2)_C$ custodial symmetry due to the presence of $T$ in their
definitions.  Notice that quartic couplings involving photons remain untouched
by the genuine quartic anomalous interactions at the order $D=4$.  The Feynman
rules for the quartic couplings generated by these operators can be found in
the last article of Ref.\ \cite{Appelquist}.

In chiral perturbation theory, the $p^4$ contributions to the processes
(\ref{wwwn})--(\ref{wwzz}) arise from the tree level insertion of $p^4$
operators, as well as from one-loop corrections due to the $p^2$ interactions,
which renormalize the $p^4$ operators \cite{Appelquist}. However, the loop
corrections to the scattering amplitudes are negligible in comparison to the
$p^4$ contributions for the range of values of the couplings and
center--of--mass energies considered in this paper. Therefore, numerically,
our analysis is consistent even though we neglected the loop corrections and
kept only the tree--level $p^4$ contributions.

In the effective--$W$ approximation \cite{hwz,evb}, the signals
(\ref{wwwn})--(\ref{wwzz}) are described by the scattering $V_L V_L \to V_L
V_L$. These processes, however, do not respect the unitarity of the
partial--wave amplitudes $(a_\ell^I)$ at large subprocess center--of--mass
energies $M_{VV}$ \cite{lee,unit,equ}. Therefore, the chiral 
expansion is valid
only for values of $M_{VV}$ and $\alpha_i$ such that $| a_\ell^I| \lesssim
1/2$. For higher $VV$ invariant masses, rescattering effects are important to
unitarize the amplitudes. Taking into account this fact, we conservatively
restricted our analyses to invariant masses $M_{VV}<$ 1.25 TeV.  This
requirement corresponds to a sharp--cutoff unitarization
\cite{unit2}.

\section{Calculational tools}
\label{tools}

In order to study the quartic couplings of vector bosons we analyzed the
processes (\ref{wwwn})--(\ref{wwzz}) which may receive contributions from
anomalous $WWZZ$ and $ZZZZ$ interactions.  The signal for vector boson fusion
in the $\gamma\gamma$ ($e\gamma$) reactions is characterized by the presence
of two central vector bosons as well as by two (one) extra ones in the forward
and backward regions of the detector, which can be used to tag the events.
Therefore, we ordered the produced vector bosons according to their rapidities
and assumed that the strongly scattered ones have smaller rapidities in
absolute value.

The signal and backgrounds were simulated at the parton level with full tree
level matrix elements.  We include in our calculations all SM and anomalous
contributions that lead to the final states (\ref{wwwn})--(\ref{wwzz}) ,
taking into account the effect of interferences between the anomalous and SM
amplitudes. This was accomplished by numerically evaluating helicity
amplitudes for all subprocesses using MADGRAPH \cite{madg} in the framework of
HELAS \cite{hellas}, with the anomalous couplings arising from the Lagrangians
(\ref{eff:4})--(\ref{eff:10}) being implemented as additional Fortran
routines. For the sake of illustration, the SM background for the processes
(\ref{wwwn}) to (\ref{wwzz}) requires the evaluation of 87, 54, 240, and 74
Feynman diagrams respectively.

In our calculations we included the reconstruction efficiency for central
$W$'s and $Z$'s which are required to have $|\eta_{W(Z)}| < 1$.  Using the
results of Ref.\ \cite{vernon}, we assumed that reconstruction efficiency of a
$W$ ($Z$) is 85\% (74\%) while the probability for misidentifying a $Z$ as a
$W$ (a $W$ as a $Z$) being 22\% (10\%). We studied the probability of tagging
forward $W$'s and $Z$'s with $1.5 < |\eta_{W(Z)}| < 3$ by consistently
including their decay into jets. Our analysis showed that just one of the jets
coming from a vector boson in the above rapidity region has $|\eta_j|<2$ and
that the tagging efficiency of the spectator $W$ from the presence of this jet
is 40\%.

The most promising mechanism to generate hard photon beams in an $e^+
e^-$ linear collider is laser backscattering, which can lead to a rich
source of $e \gamma$ and $\gamma \gamma$ interactions with essentially
the same luminosity and center-of-mass energy of the parent $e^+e^-$
collider \cite{telnov}.  We verified that the polarization of the
beams do not change significantly our results, therefore, we present
our results for unpolarized electron and laser beams for the sake of
simplicity. In this case, the backscattered photon distribution
function \cite{laser} is
\begin{equation}
F_{\gamma/e}  (x,\xi) \equiv \frac{1}{\sigma_c} \frac{d\sigma_c}{dx} =
\frac{1}{D(\xi)} \left[ 1 - x + \frac{1}{1-x} - \frac{4x}{\xi (1-x)} +
\frac{4
x^2}{\xi^2 (1-x)^2}  \right] \; ,
\label{f:l}
\end{equation}
with
\begin{equation}
D(\xi) = \left(1 - \frac{4}{\xi} - \frac{8}{\xi^2}  \right) \ln (1 + \xi) +
\frac{1}{2} + \frac{8}{\xi} - \frac{1}{2(1 + \xi)^2} \; ,
\end{equation}
where $\sigma_c$ is the Compton cross section, $\xi \simeq 4
E\omega_0/m_e^2$, $m_e$ and $E$ are the electron mass and energy
respectively, and $\omega_0$ is the laser-photon energy. The quantity
$x$ stands for the ratio between the scattered photon and initial
electron energy and its maximum value is     
\begin{equation}
x_{\text{max}}= \frac{\xi}{1+\xi} \; .
\end{equation}
In what follows, we assumed that the laser frequency is such that $\xi = 2(1
+\sqrt{2})$, which leads to the hardest possible spectrum of photons with a
large luminosity since the creation of soft $e^+e^-$ pairs in hard
$\gamma$--laser interaction is avoided with this choice. In this case,
$x_{\text{max}}\simeq 0.83 $.

The cross section for $\gamma\gamma$ fusion processes can be obtained by
folding the elementary cross section for the $\gamma \gamma$ subprocesses
with the photon distributions, {\it i.e.},
\begin{equation}
d\sigma (e^+e^-\rightarrow \gamma\gamma \rightarrow X)(s) =
\int dx_1 dx_2~ F_{\gamma/e}(x_1,\xi)~F_{\gamma/e}(x_2,\xi)~ 
d \hat\sigma  (\gamma\gamma \rightarrow X) (\hat s) \; ,
\end{equation}
where $\sqrt{s}$ ($\sqrt{\hat{s}}$) is the $e^+e^-$ ($\gamma\gamma$)
center-of-mass energy. In the case of $e\gamma$ collisions we should drop one
of the integrals on the photon spectrum. In our analyses we considered two
$e^+e^-$ center--of--mass energies $\sqrt{s} = 2$ and 3 TeV which leads to a
maximum $\gamma\gamma$ ($e\gamma$) center--of--mass energy of 1.7 and 2.5 (1.8
and 2.7) TeV respectively. We assumed that the integrated luminosity for the
parent $e^+e^-$ machine is 500 fb$^{-1}$ which is a conservative choice since
we can always tune up the beam shape in order to boost the $\gamma\gamma$ and
$e\gamma$ luminosities \cite{telnov}.

\section{Signal and background properties}
\label{prop}

Strongly interacting symmetry breaking sectors (SEWS) modify the dynamics of
longitudinal vector bosons. However, it is impossible to determine the
polarization of vector bosons on an event--by--event basis, and consequently,
we have to work harder to extract the SEWS signal. Taking into account that
the electroweak production of transversely polarized vector bosons in the SM
is approximately independent of the Higgs boson mass, and that the $V_L V_L$
production is small for light Higgs bosons \cite{bagger}, we define the signal
for SEWS as an excess of events in the $V V$ scattering channels with respect
to the SM with a light Higgs, {\em i.e.}
\begin{equation}
\sigma_{\text{signal}} \equiv \sigma (\alpha_i)
-  \sigma^{\text{lh}}_{\text{sm}}  \;\;\; ,
\label{sig:def}
\end{equation}
where $\sigma^{\text{lh}}_{\text{sm}} = \sigma_{\text{sm}}\Bigr |_{M_H =100
\text{ GeV}}$ and we sum over the vector-boson polarizations.  In principle,
we can have a signal even for $\alpha_i\equiv 0$, indicating the existence of
SEWS, since there is no Higgs in our model to cut off the growth of the
scattering amplitudes.  In this case that it is possible to establish the
existence of SEWS, we should verify whether the anomalous couplings $\alpha_i$
are compatible with zero or not. In this scenario we define the
$\sigma_{\text{signal}}$ with respect to vanishing $\alpha$'s, {\em i.e.}
\begin{equation}
\sigma_{\text{signal}} \equiv \sigma (\alpha_i) -
\sigma (0) \; .
\label{sig:def2}
\end{equation}

The most general expression for the total cross sections of the processes
(\ref{wwwn})--(\ref{wwzz}) can be written as
\begin{equation}
\sigma \equiv \sigma^{hh}_{\text{sm}} + \alpha_j~
\sigma^{\alpha_j}_{\text{int}} + \alpha_i \alpha_j~ \sigma^{\alpha_i
\alpha_j}_{\text{ano}} \; ,
\label{crosssection}
\end{equation}
where $\sigma^{hh}_{\text{sm}}$, $\sigma^{\alpha_j}_{\text{int}}$, and
$\sigma^{\alpha_i \alpha_j}_{\text{ano}}$ are, respectively, the SM cross
section without the inclusion of the Higgs boson effects, interference between
the heavy Higgs SM and the anomalous contributions and the pure anomalous
cross section.  $\alpha_i$ stands for any of the anomalous quartic couplings
appearing in Eqs.\ (\ref{eff:4})--(\ref{eff:10}).

One of the important features of the vector boson scattering is the presence
of vector bosons at large rapidities which can be used to tag the events. In
our analyses we required the existence of spectator vector bosons with
\begin{equation}
1.5 < |\eta_{W(Z)}| < 3 \; ,
\end{equation}
which lead to a detectable spectator jet 40\% of the time. Since, we detect
only one of the jets coming from forward $W$'s and $Z$'s, there is one
additional background due to the production of top quark pairs when the $b$
from the top decay is taken as the tagging forward jet. Therefore we vetoed
these events tagging $b$ jets, assuming a $b$ tagging efficiency of
60\%. Moreover, we assume a probability of 5\% that a light-quark jet is
mistagged as a $b$--quark jet.


\subsection{$e\gamma$ mode}

\begin{figure}
\begin{center}
\mbox{\epsfig{file=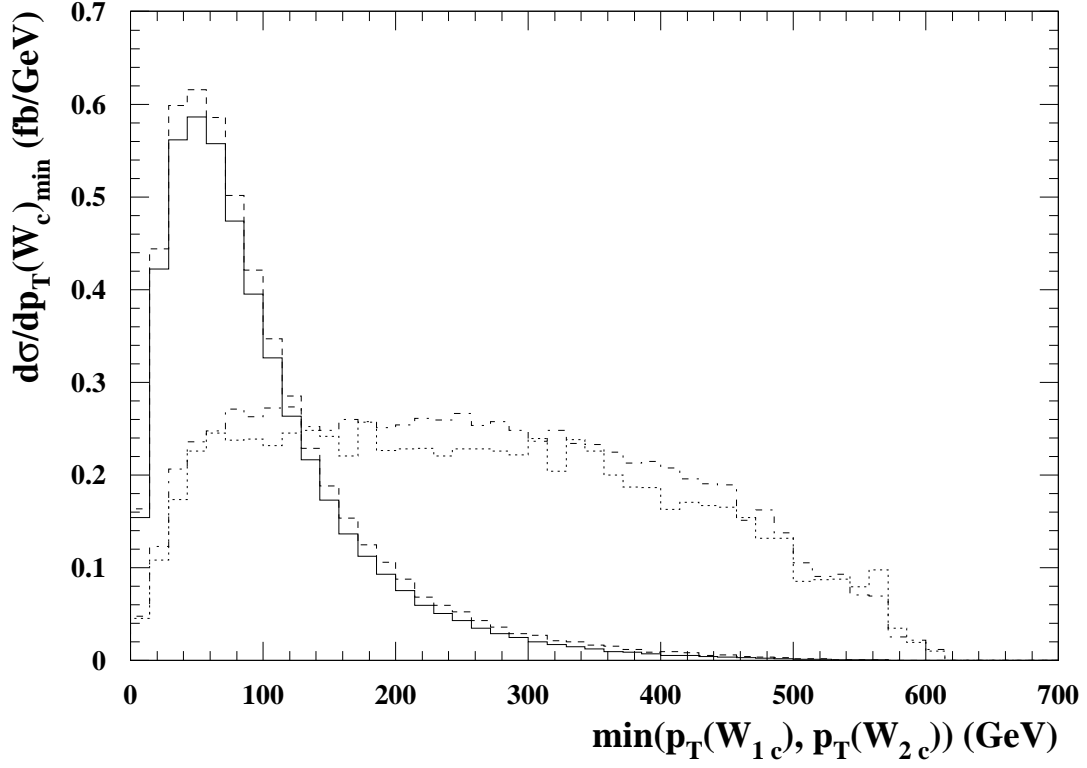,width=1\textwidth}}
\end{center}
\caption{Minimal transverse momentum of the central $W$ bosons
for the process $e^+ e^- \to e^- \gamma \to W^- W^+ W^- \nu_e$ at
$\sqrt{s_{ee}} = 2$ TeV.  The solid (dashed) line stands for the SM with a
light Higgs (heavy Higgs) and the dotted (dotted-dashed) is the anomalous
contribution for $\alpha_{4(5)}=0.05$.}
\label{fig:pt_eaw}
\end{figure}

\begin{figure}
\begin{center}
\mbox{\epsfig{file=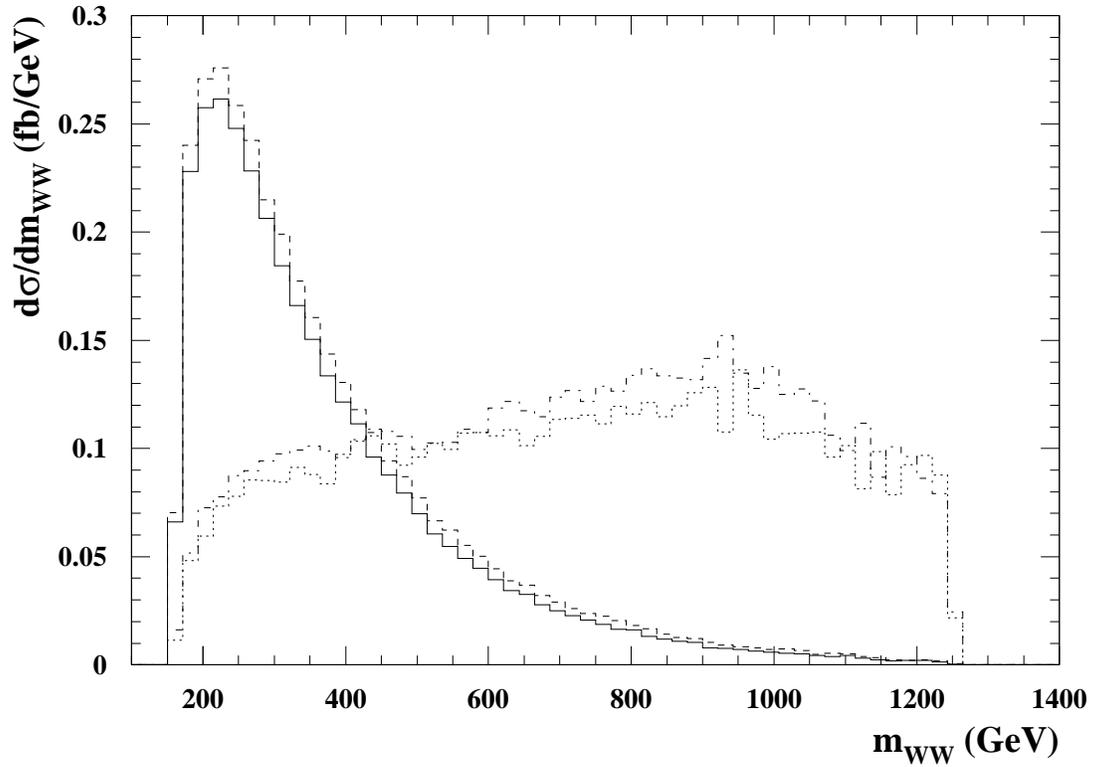,width=1\textwidth}}
\end{center}
\caption{ 
  Invariant mass distribution of the central $W$ boson pair for the process
  $e^+ e^- \to e^- \gamma \to W^- W^+ W^- \nu_e$ at $\sqrt{s_{ee}} = 2$
  TeV. We applied the regularization cut $m_{WW} < 1.25$ TeV and employed the
  conventions as in Fig.\ \protect\ref{fig:pt_eaw}.  }
\label{fig:mww_eaw}
\end{figure}

In order to suppress the backgrounds and enhance the signal for the anomalous
quartic interactions in processes (\ref{wwwn}) and (\ref{wzzn}) we applied the
following set of cuts:
\begin{itemize}

\item [$(i)$] We required the tagging of a large rapidity vector boson as
described above.

\item [$(ii)$] We vetoed events presenting $b$ tagged jets in order to reduce
the top production background.

\item [$(iii)$] We demanded the presence of a pair $W^+W^-$ ($ZZ$) in the
process $e \gamma \to W^- W^+ W^- \nu_e$ ($W^- Z Z \nu_e$) with $p_T^{W(Z)} >
200$ GeV and $|\eta_{W(Z)}| < 1$. In order to access the relevance of this
$p_T$ cut see Fig.\ \ref{fig:pt_eaw}.

\item [$(iv)$] We also required the invariant mass of the vector boson pair to
be in the range $0.5 < M_{VV} < 1.25$ TeV. The upper limit of this
cut is quite important since it prevents the effective operators
(\ref{eff:4})--(\ref{eff:10}) to be used in a energy regime where unitarity is
violated and rescattering effects become important. The lower limit of this
cut aims to reduce the backgrounds; see Fig.\ \ref{fig:mww_eaw}.

\end{itemize}

We display in Table \ref{sig:ea} our results for the coefficients $\sigma$ in
Eq.\ (\ref{crosssection}) after applying the above cuts, however before taking
into account the detection efficiencies. In the $W^-ZZ\nu_e$ production, the
central gauge boson pair can be either $W^-Z$ or $ZZ$, therefore, we show
these two cases in Table \ref{sig:ea}. As expected, the cross sections rise as
the center--of--mass energy increases. Moreover, it is clear from this table
that the $W^-W^+W^-\nu_e$ process not only leads to a larger statistics but
also it is more sensitive to the quartic anomalous couplings. In order to
obtain the signal cross section, we must fold the results in Table
\ref{sig:ea} with the reconstruction efficiencies and take into account the
misidentification probabilities since both processes, $W^-W^+W^-\nu_e$ and
$W^-ZZ\nu_e$, can lead to events $WW$ + jet and $ZZ$ + jet, where the pair of
gauge bosons is central and the single jet comes from a forward $W$ or $Z$.
For instance, about 5\% of the $ZZ$ + jet signal events are due to vector
boson misidentification in the reaction $WWW\nu_e$ for $\alpha_4 = \alpha_5 =
0$.

Table \ref{bkg:ea} contains the total light Higgs background cross section for
processes (\ref{wwwn}) and (\ref{wzzn}) after cuts and detection efficiencies.
We also display in this table the fraction $F$ of the total background due to
each reaction. Notice that the bulk of the background to the $WW$ + jet events
is due to the production of $WWW\nu$ in the scope of the SM. However, some
fraction of the background ($\simeq 9$\%) to the $ZZ$ + jet events is due to
misidentification of $W$'s generated by reaction (\ref{wwwn}). In $e\gamma$
production the $t\bar{t}$ background is not important.

\subsection{$\gamma\gamma$ mode}

The production of $W^-W^+W^-W^+$ and $W^-ZZW^+$ in $\gamma\gamma$ collisions
due to anomalous quartic interactions can be enhanced using the same cuts
employed in the $e\gamma$ case, with the exception that we now require the
tagging of two forward $W$'s.  Table \ref{sig:aa} contains our results for the
parameters appearing in (\ref{crosssection}) after cuts but before introducing
the detection efficiencies. In this mode, the $W^-W^+W^-W^+$ production not
only possesses the highest cross section but also it exhibits a stronger
dependence on the anomalous couplings. Analogously to the $e\gamma$ case, we
have to conveniently introduce the detection efficiencies, branching ratios,
and misidentification probabilities in order to obtain the expected number of
events.

We show the total light Higgs backgrounds in the $\gamma\gamma$ mode and its
composition in Table \ref{bkg:aa}. The $WW + 2$--jet signal events have as
major backgrounds the production of longitudinal $WWWW$ in the scope of the SM
and the $t\bar{t}$ production, with a small contribution from the
misidentification of vector bosons. It is interesting to notice that the top
pair background represents almost half of the total background at 2 TeV with
its importance decreasing at higher energies. For the $ZZ+2$--jet events, the
major background is the production of $W^-ZZW^+$ within the SM with $t\bar{t}$
production and the particle misidentification being responsible for 17--25\%
of the background events.

\section{Discussion and Conclusions}
\label{resu} 

\begin{figure}
\begin{center}
\mbox{\epsfig{file=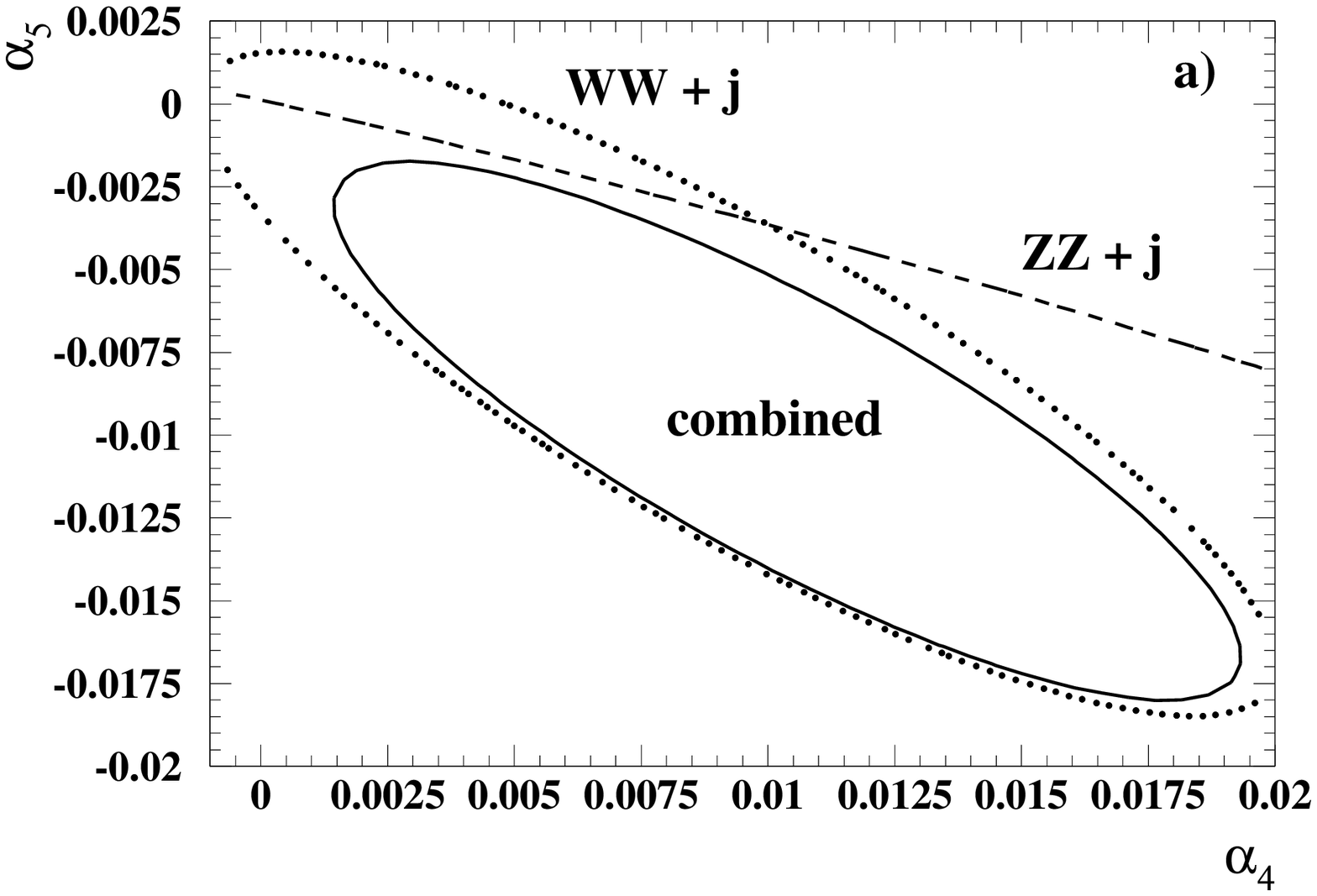,width=0.45\textwidth}}
\hfill
\mbox{\epsfig{file=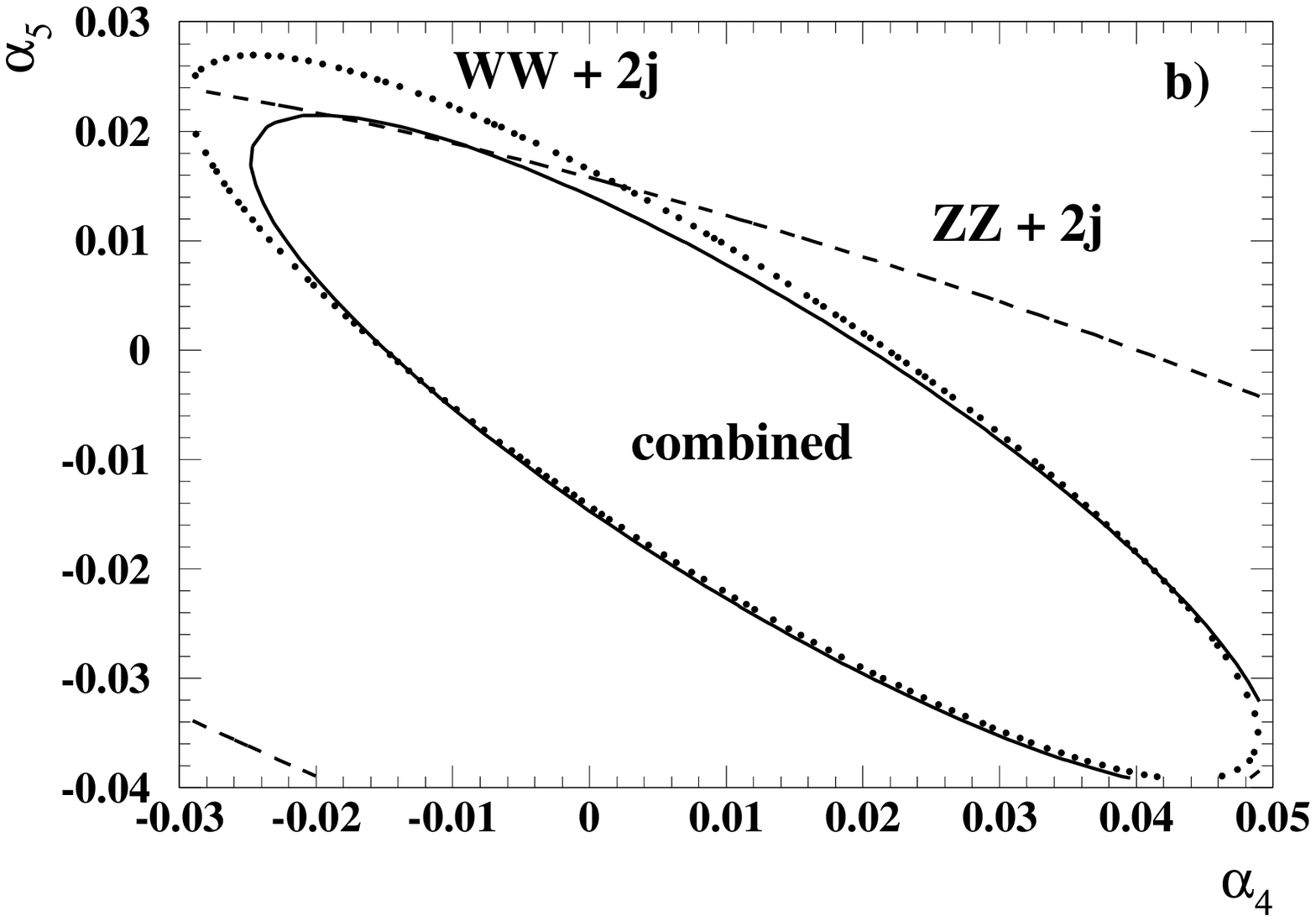,width=0.45\textwidth}}
\end{center}
\caption{ $5\sigma$ allowed region in the $(\alpha_4, \alpha_5)$ plane
  assuming that we observed the expected number of events predict by the SM
  with a light Higgs boson in the $WW$ + jet (dots), $Z Z$ + jet (dashes), and
  combined reactions (solid). We assumed a $e^+e^-$ center--of--mass energy of
  2 TeV and an integrated luminosity of 500 fb$^{-1}$. In (a) we display the
  results for a $e \gamma$ collider while (b) contains the results for the
  $\gamma\gamma$ mode.}
\label{fig:lh}
\end{figure}

Initially we studied whether an $e\gamma$ or $\gamma\gamma$ collider can
unravel the existence of a symmetry breaking dynamics different from the one
predicted by the SM. In order to do that we analyzed the reactions
(\ref{wwwn}) to (\ref{wwzz}) assuming that the number of observed events is
the one predict by the SM with a light Higgs boson. We display in Fig.\ 
\ref{fig:lh}a the $5\sigma$ allowed region in the $e\gamma$ mode for an
$e^+e^-$ center--of--mass energy of 2 TeV and an integrated luminosity of 500
fb$^{-1}$. As expected, most of $WW$ + jet events are due to $W^- W^+ W^-
\nu_e$ production and this channel leads to more stringent limits and
dominates the combined bounds. This figure clearly shows that the combined
results for the $WW$ + jet and $ZZ$ + jet events will be able to establish the
existence of a new dynamics in the symmetry breaking system since the
agreement with experiment is only possible for non-vanishing quartic
couplings. Fig.\ \ref{fig:lh}b contains the results for a $\gamma \gamma$
collider assuming the same center--of--mass energy, luminosity and confidence
level. The $\gamma \gamma$ mode at 2 TeV will not be able to establish a
departure from the SM, and we verified that the signal of new dynamics only
appears at higher center--of--mass energies, {\em e.g.} 3 TeV. For the sake of
comparison, we estimated that establishing the existence of SEWS in the LC
$e^+e^-$ mode will require an integrated luminosity of the order of 50
fb$^{-1}$ for a center--of--mass energy of 1.6 TeV.

\begin{figure}
\begin{center}
\mbox{\epsfig{file=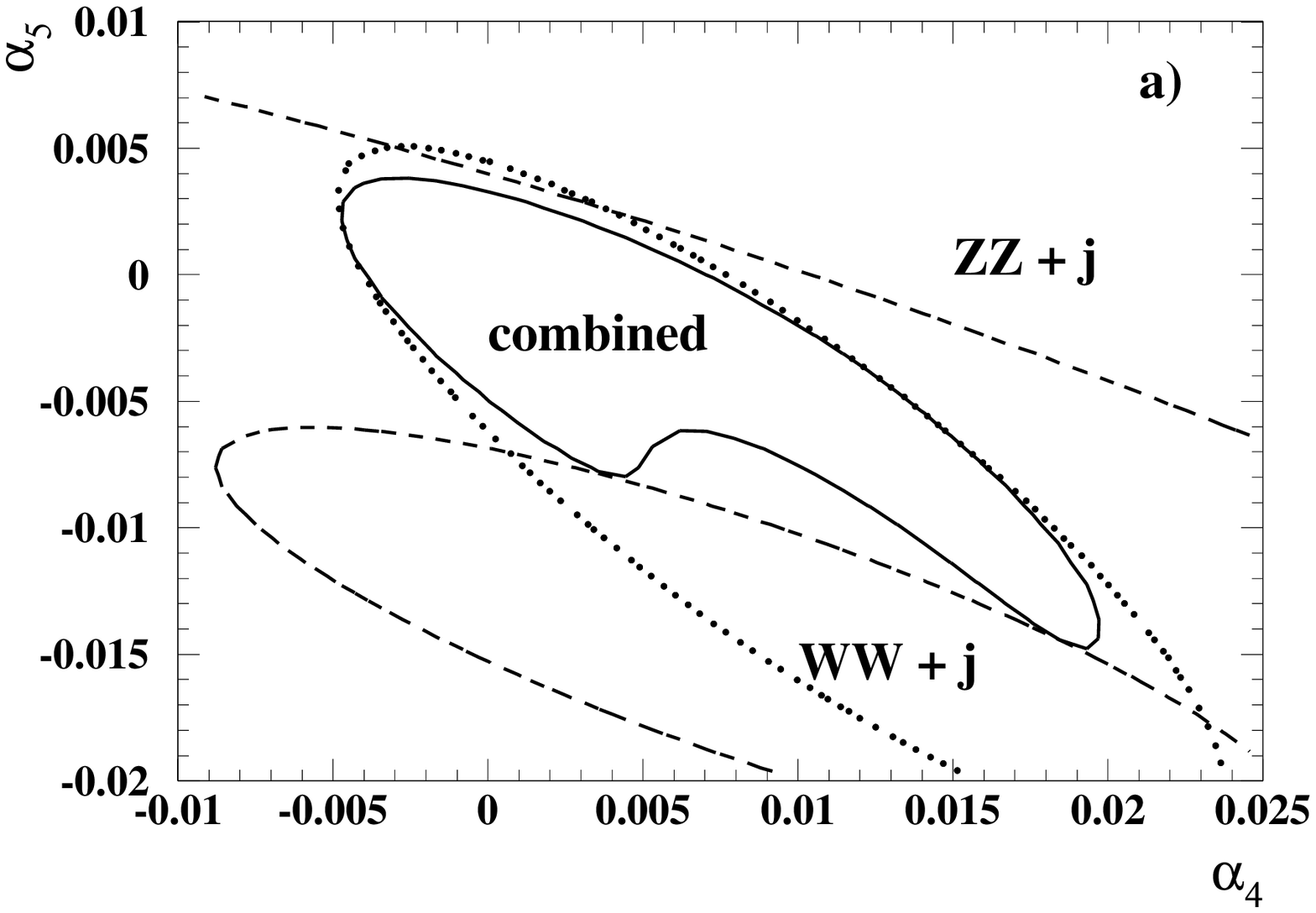,width=0.45\textwidth}}
\hfill
\mbox{\epsfig{file=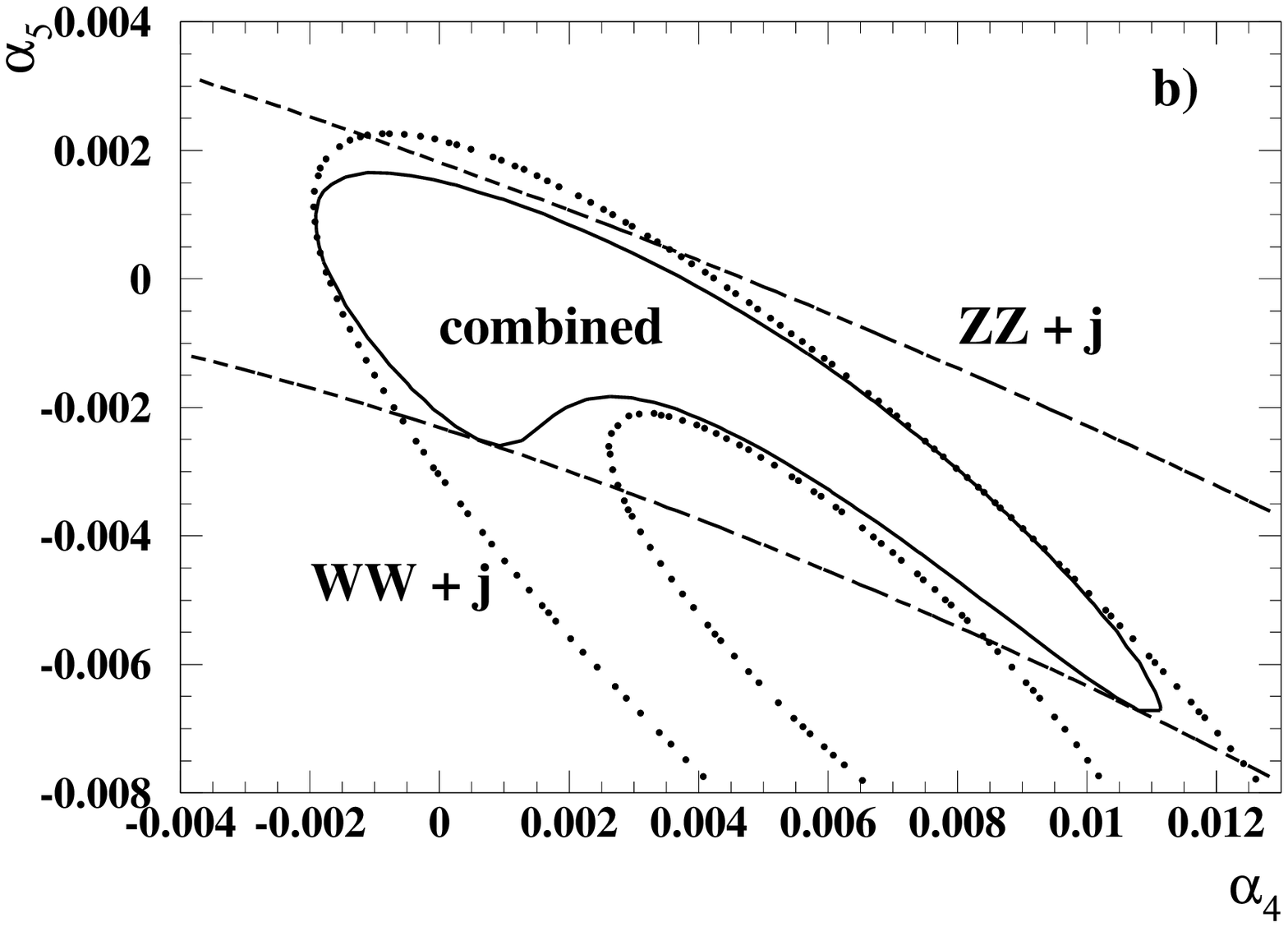,width=0.45\textwidth}}
\end{center}
\caption{ 90\% CL  allowed region in the $(\alpha_4, \alpha_5)$ plane
  assuming that we observed the number of events predict by the SM without a
  Higgs boson in the $e\gamma \to WW$ + jet (dots), $e \gamma \to Z Z$ + jet
  (dashes), and combined reactions (solid). We assumed a $e^+e^-$
  center--of--mass energy of 2 TeV (a) [3 TeV (b)] and an integrated
  luminosity of 500 fb$^{-1}$. }
\label{fig:eahh}
\end{figure}

Having established the existence of a new dynamics we probed the LC capability
to constrain anomalous quartic vector boson couplings by assuming that the
number of observed events is the one predicted by the SM without the Higgs
boson; see Eq.\ (\ref{sig:def2}). We present in Fig.\ \ref{fig:eahh} the 90\%
CL bounds on the $SU(2)_C$ conserving quartic couplings coming from the
reactions (\ref{wwwn}) and (\ref{wzzn}), assuming center--of--mass energies of
2 and 3 TeV and an integrated luminosity of 500 fb$^{-1}$. Here the best
limits also come from the $WW$ + jet production, however, the combined results
are much more restrictive than this channel alone since the $ZZ$ + jet allowed
region has a different orientation than the $WW$ + jet one. Moreover, the
constraints improve by a factor of ${\cal O}(2)$ when the center--of--mass
energy increases from 2 TeV to 3 TeV.

The bounds on the couplings $\alpha_4$ and $\alpha_5$ obtained in the $e
\gamma$ mode of the LC are of the same order of the ones coming from the $e^+
e^-$ mode \cite{bhk}. Moreover, the $e \gamma$ mode possesses only one allowed
region in the plane $(\alpha_4, \alpha_5)$, while the reactions $e^+ e^- \to
\bar{\nu} \nu W^+ W^-$ and and $\bar{\nu} \nu Z Z$ lead to two allowed
regions, one in the vicinity of $(0,0)$ and the second one having only
non-vanishing values of the anomalous couplings; see Ref.\ \cite{bhk}.
Therefore, the information gathered in the $e \gamma$ mode can be used to
reduced the ambiguities of the $e^+e^-$ mode, analogously to what happens in
the $e^-e^-$ mode \cite{bhk}.

\begin{figure}
\begin{center}
\mbox{\epsfig{file=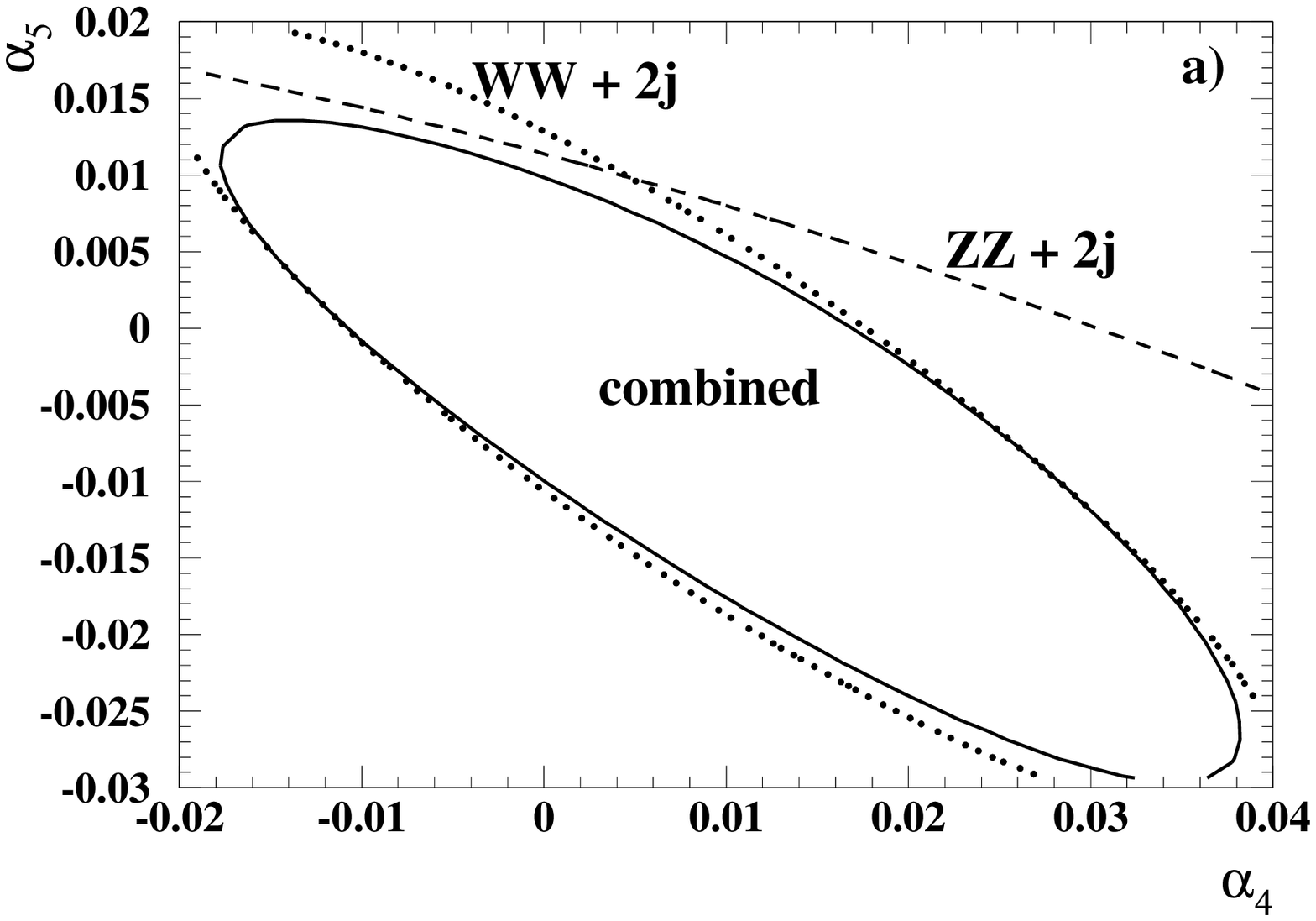,width=0.45\textwidth}}
\hfill
\mbox{\epsfig{file=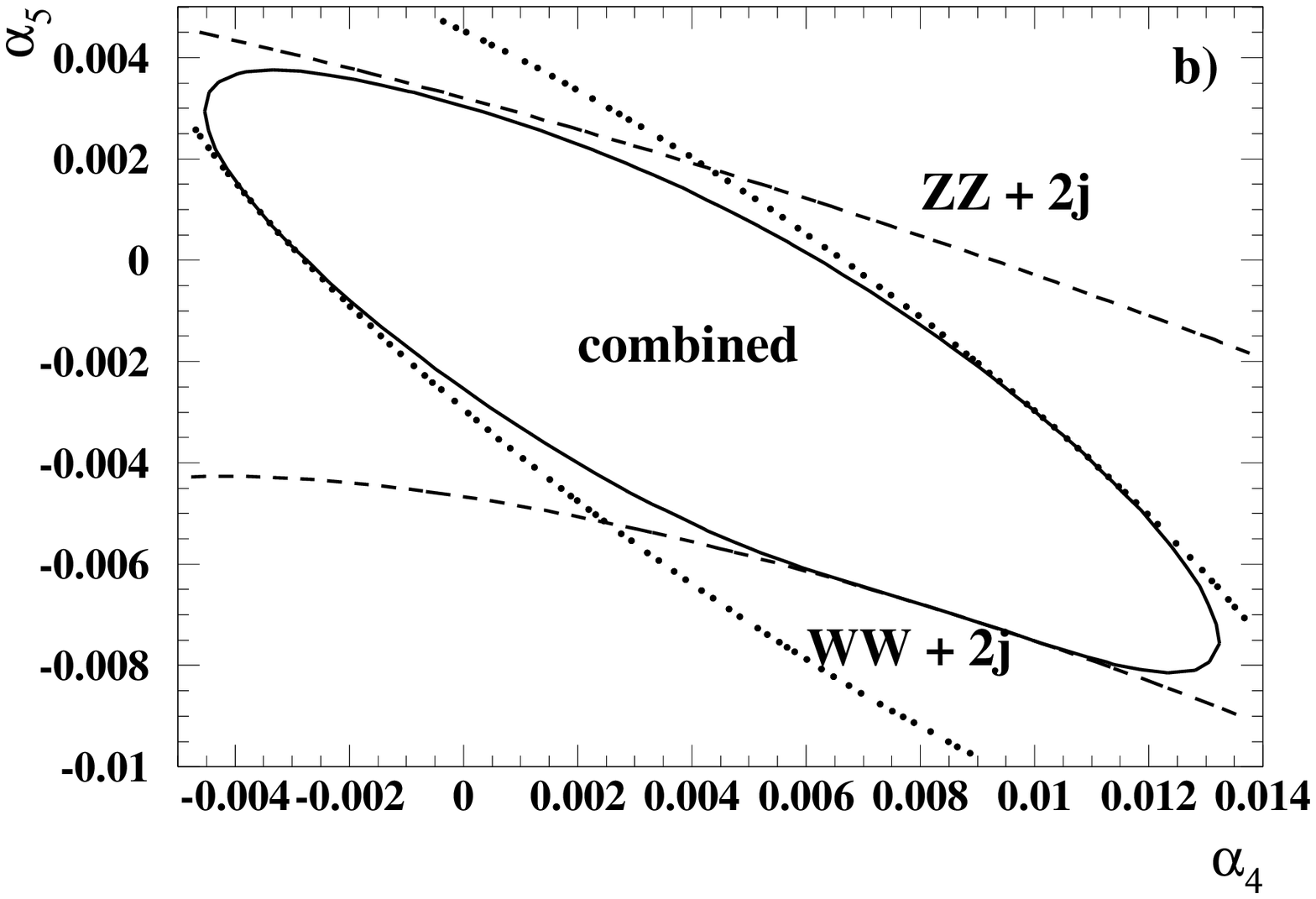,width=0.45\textwidth}}
\end{center}
\caption{ 90\% CL  allowed region in the $(\alpha_4, \alpha_5)$ plane
  assuming that we observed the number of events predict by the SM without a
  Higgs boson in the $\gamma \gamma \to WW$ + 2 jets (dots), $\gamma \gamma
  \to Z Z $ + 2 jets (dashes), and combined reactions (solid). We assumed a
  $e^+e^-$ center--of--mass energy of 2 TeV (a) [3 TeV (b)] and an integrated
  luminosity of 500 fb$^{-1}$. }
\label{fig:aahh}
\end{figure}

Fig.\ \ref{fig:aahh} shows the attainable bounds on $\alpha_4$ and $\alpha_5$
through the reactions $\gamma \gamma \to WW$ + 2 jets and $Z Z$ + 2 jets for
an integrated luminosity of 500 fb$^{-1}$ and center--of--mass energies of 2
and 3 TeV. In this mode, the best limits originate from the $WW$ + 2--jet
events, which receives most of contributions from $W^+W^-W^-W^+$, since this
process exhibits the largest cross section. Moreover, the process $Z Z$ + 2
jets turns out to be important to exclude a large fraction of the region
allowed by the $WW$ + 2 jet production. Like the $e \gamma$ mode, the $\gamma
\gamma$ collider also gives rise to just one allowed region around the origin,
therefore it is complementary to the $e^+e^-$ mode too. However, the limits
that can be obtained from this mode are weaker than the ones coming the the
$e^+e^-$ and $e \gamma$ modes.

\begin{figure}
\begin{center}
\mbox{\epsfig{file=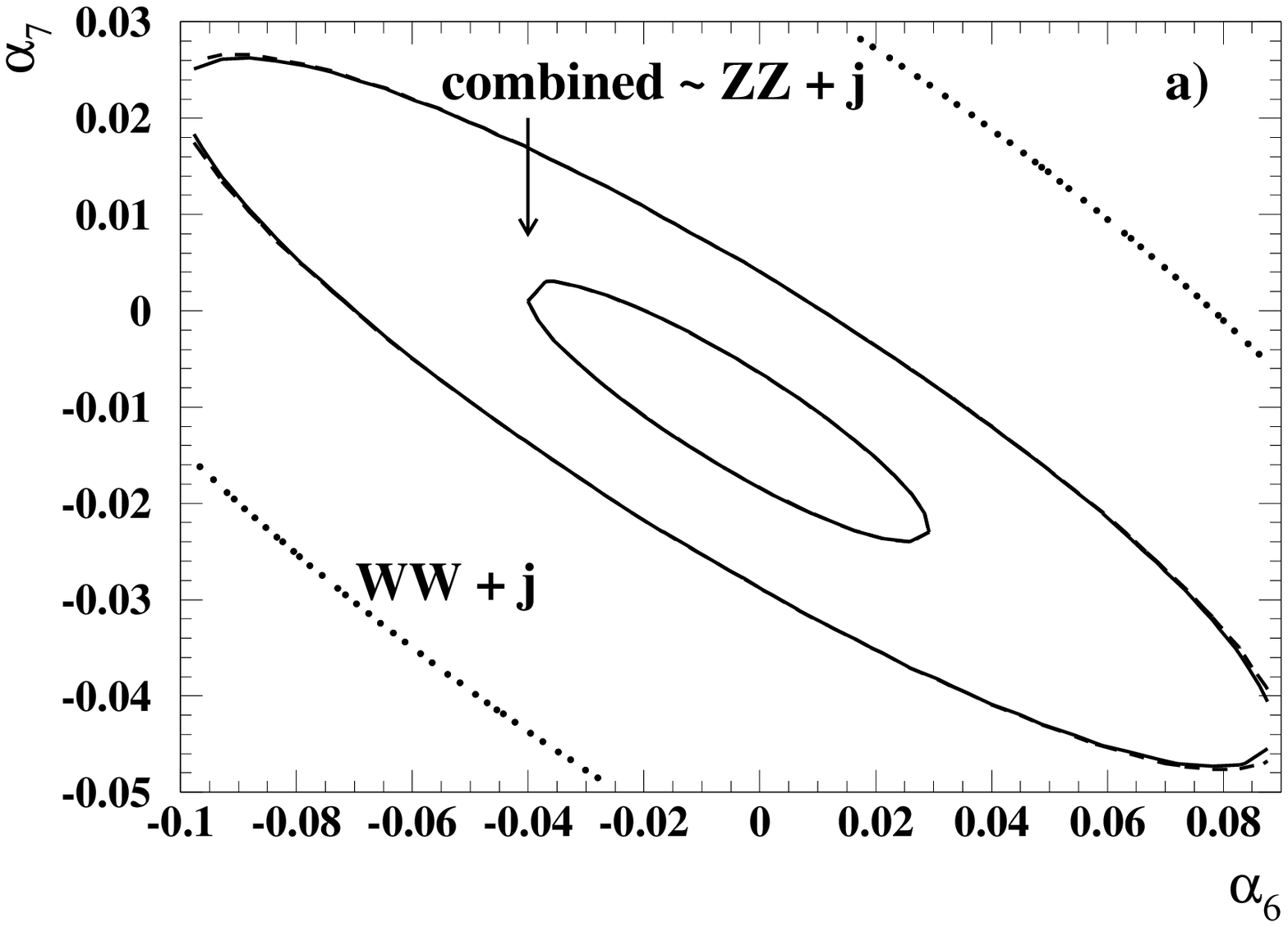,width=0.45\textwidth}}
\hfill
\mbox{\epsfig{file=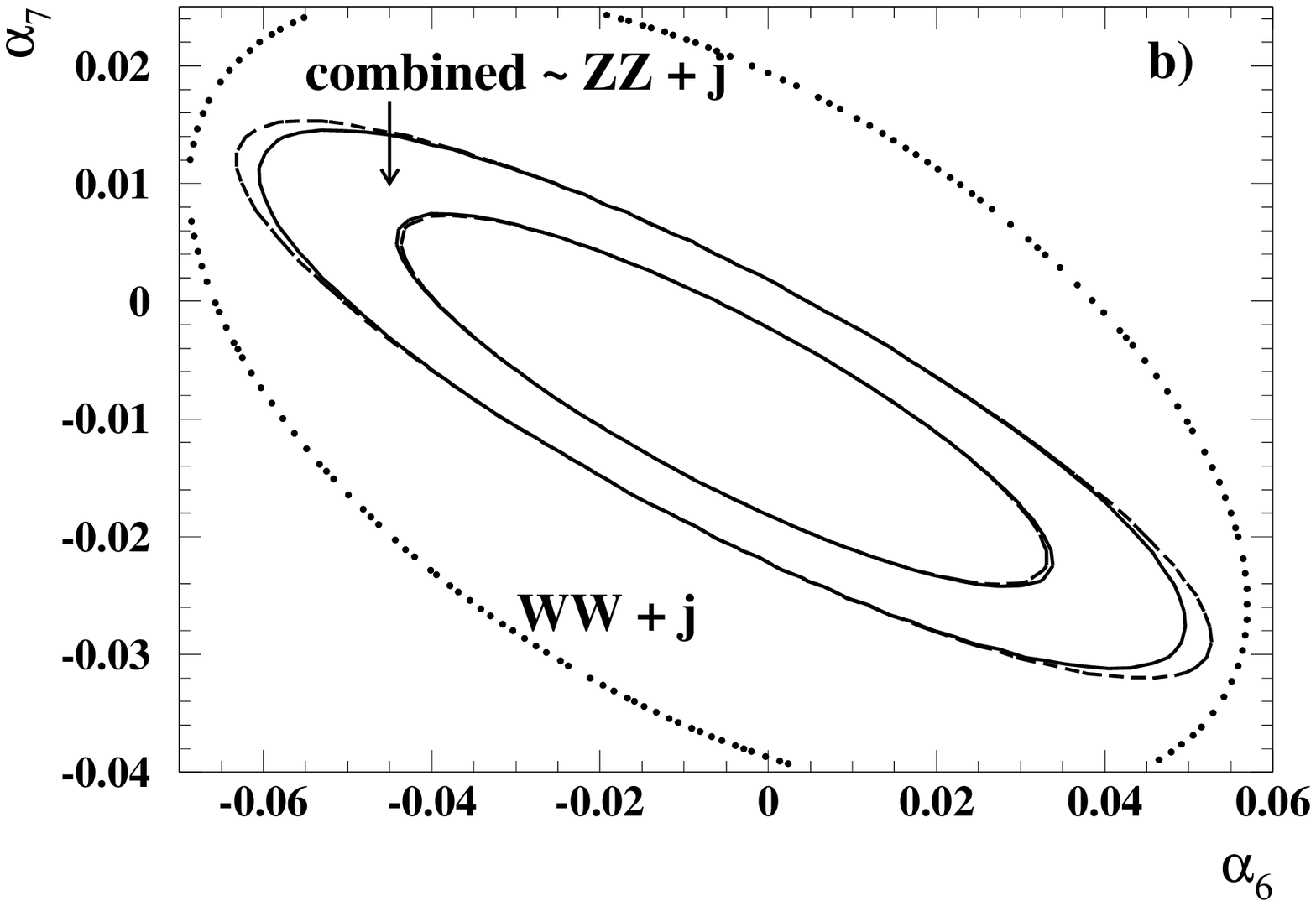,width=0.45\textwidth}}
\hfill
\mbox{\epsfig{file=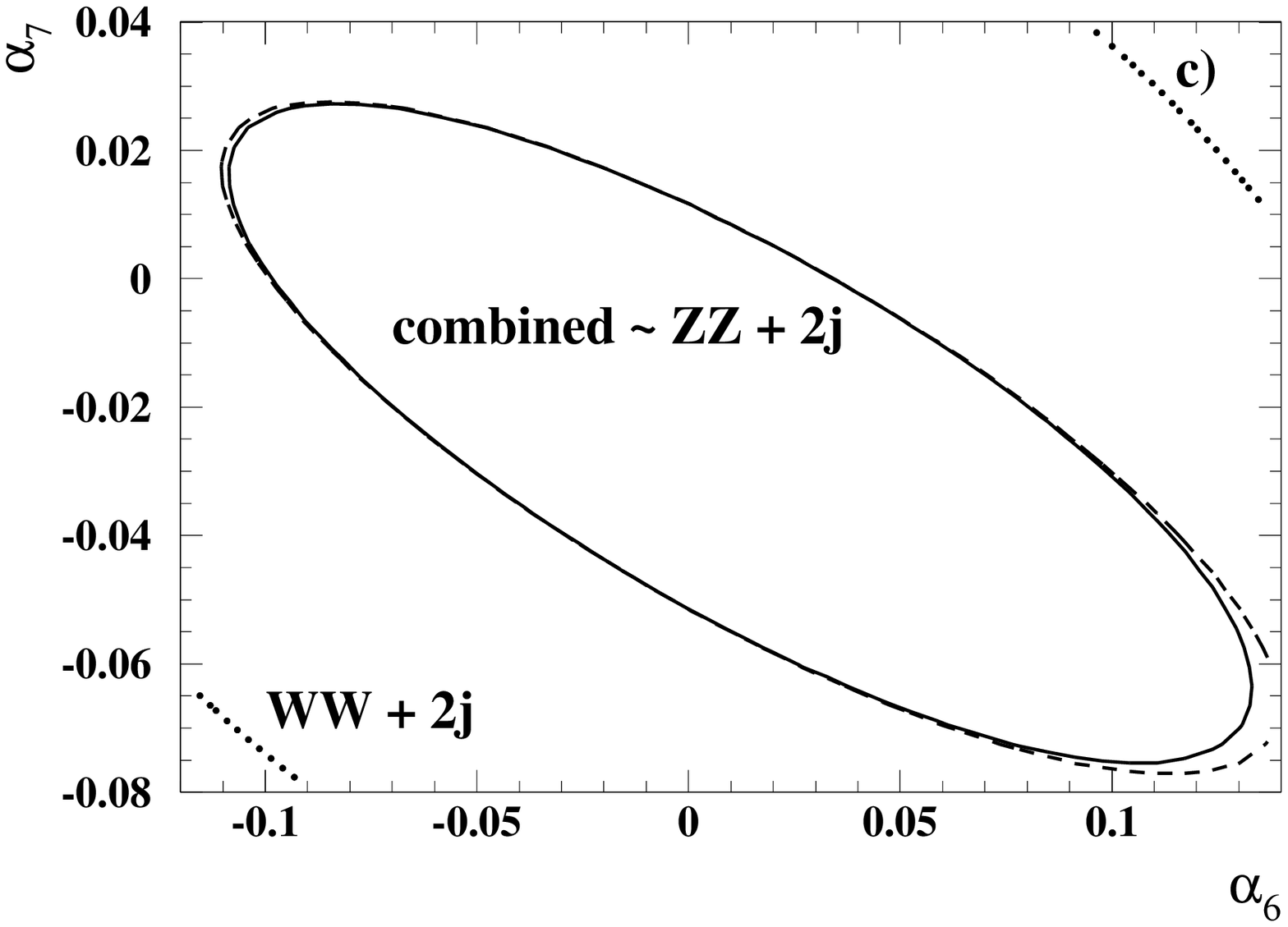,width=0.45\textwidth}}
\hfill
\mbox{\epsfig{file=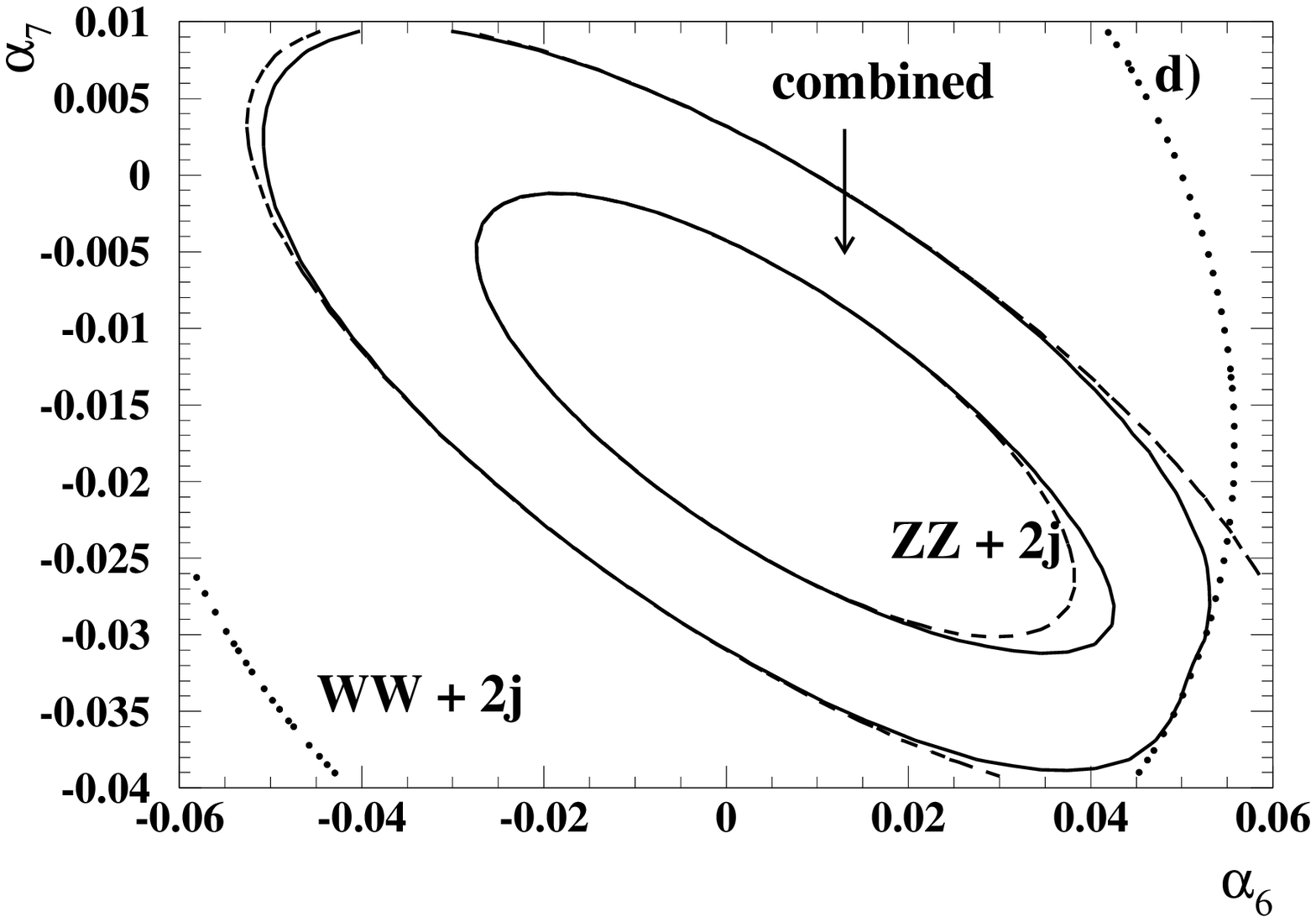,width=0.45\textwidth}}
\end{center}
\caption{ Combined 90\% CL  allowed region in the $(\alpha_6, \alpha_7)$ plane
  assuming that we observed the number of events predict by the SM without a
  Higgs boson in the $e \gamma$ mode [(a) and (b)] and $\gamma \gamma$ one
  [(c) and (d)].  We assumed a $e^+e^-$ center--of--mass energy of 2 TeV in
  (a) and (c) and 3 TeV in (b) and (d), and an integrated luminosity of 500
  fb$^{-1}$. }
\label{fig:67}
\end{figure}

Up to now we concentrated our analyses on the $SU(2)_C$ conserving quartic
operators (\ref{eff:4}) and (\ref{eff:5}). First of all, the effective
interaction (\ref{eff:10}) can not be constrained by any of the processes
studied here since it leads only to a $Z Z Z Z$ vertex.  The vertex $W^+ W^- Z
Z$ associated to $\alpha_6$ ($\alpha_7$) is equal to the one generated by
$\alpha_4$ ($\alpha_5$), however the operators (\ref{eff:4}) and (\ref{eff:5})
do not induce $WWWW$ interactions. Therefore, $\alpha_6$ and $\alpha_7$ do not
contribute to the most stringent processes $e \gamma \to W^+ W^- W^+ \nu_e$
and $\gamma \gamma \to W^+ W^- W^+ W^-$.  From this fact we expect that the
bounds on these operators should be much weaker, analogously to what takes
place in the $e^+e^-$ mode. In Fig.\ \ref{fig:67} we display the combined
limits on these operators for a 2 and 3 TeV LC.

In brief, the $e \gamma$ can help us to have a better understanding of the
symmetry breaking sector of the electroweak interactions since it leads to
constraints ${\cal O}(0.005)$ on the $SU(2)_C$ conserving quartic gauge boson
interactions for a 2 TeV LC.  Moreover, this mode is complementary to the
$e^+e^-$ one since it allow us to resolve the ambiguity that the last mode
presents on the limits on these couplings. Analogously to the traditional
$e^+e^-$ mode the $e\gamma$ bounds on $SU(2)_C$ violating operators are also
much less stringent.  We also showed that the $\gamma \gamma$ mode is
complementary to the $e^+ e^-$ one, however that mode leads to bounds that are
a factor 5--10 weaker.


\acknowledgments

This work was supported by Conselho Nacional de Desenvolvimento
Cient\'{\i}fico e Tecnol\'ogico (CNPq), by Funda\c{c}\~ao de Amparo \`a
Pesquisa do Estado de S\~ao Paulo (FAPESP), by Programa de Apoio a N\'ucleos
de Excel\^encia (PRONEX), and by the U.~S.~Department of Energy under contract
number DE-FG03-94ER40833.





%
\begin{table}[hbt]
\begin{center}
\begin{tabular}{cccccccc}
reaction  
& $\sqrt{s_{ee}}$ TeV 
& $\sigma^{hh}_{\text{sm}}$ (fb) 
& $\sigma_{\text{int}}^{\alpha_4}$ (fb)   
& $\sigma_{\text{ano}}^{\alpha_4\alpha_4}$ (fb)
& $\sigma_{\text{int}}^{\alpha_5}$ (fb)
& $\sigma_{\text{ano}}^{\alpha_5\alpha_5}$ (fb)
& $\sigma_{\text{ano}}^{\alpha_4\alpha_5}$ (fb)
\\ 
\hline
$W^- [W^+ W^-] \nu_e$
& 2
& 4.84
& -74.3
& 19500.
& 31.0
& 21500.
& 34900.
\\
$W^- [W^+ W^-] \nu_e$
& 3
& 10.3
& -319.
& 121$\times 10^3$
& 99.
& 133$\times 10^3$
& 221$\times 10^3$
\\
$W^- [Z Z] \nu_e$
& 2
& 1.07
& 24.4
& 404.
& 67.6
& 2716
& 1901
\\
$W^- [Z Z] \nu_e$
& 3
& 2.83
& 95.9
& 2030
& 258.
& 12600.
& 8930.
\\
$[W^- Z] Z \nu_e$
& 2
& 0.256
& 3.82
& 128.
& 8.26
& 383.
& 248.
\\
$[W^- Z] Z \nu_e$
& 3
& 0.38
& 9.25
& 747.
& 15.0
& 977.
& 63.0
\end{tabular}
\end{center}
\caption{Values for the standard model, pure anomalous and interference cross
sections after cuts, according to Eq.\ (\protect\ref{crosssection}), for $e
\gamma \to W^-W^+W^-\nu_e$ and $W^- Z Z \nu_e$ and several $e^+e^-$
center--of--mass energies.  We display between brackets the vector bosons
produced in the central region.}
\label{sig:ea}
\end{table}

%
\begin{table}[hbt]
\begin{center}
\begin{tabular}{ccccccc}
events
& $\sqrt{s_{ee}}$ TeV 
& $\sigma^{lh}_{\text{sm}}$ (fb) 
& $F (WWW\nu)$
& $F (W[ZZ]\nu)$
& $F (Z[WZ]\nu)$
& $F(t \bar{t})$
\\ 
\hline
$WW$ + jet
& 2
& 0.480
& 96.2\%
& 1.0\%
& 1.5\%
& 1.2\%
\\
$WW$ + jet
& 3
& 0.879
& 96.9\%
& 1.2\%
& 1.1\%
& 0.7\%
\\
$ZZ$ + jet
& 2
& 6.5$\times 10^{-2}$
& 9.2\%
& 86.5\%
& 4.6\%
& 0.0\%
\\
$ZZ$ + jet
& 3
& 0.139
& 8.6\%
& 88.5\%
& 2.9\%
& 0.0\%
\end{tabular}
\end{center}
\caption{Total light Higgs background cross section in the $e\gamma$ mode after
 cuts and detection efficiencies as well as its composition. We display
 between brackets the vector bosons produced or identified in the central
 region.}
\label{bkg:ea}
\end{table}

%
\begin{table}[hbt]
\begin{center}
\begin{tabular}{cccccccc}
reaction  
& $\sqrt{s_{ee}}$ TeV 
& $\sigma^{hh}_{\text{sm}}$ (fb) 
& $\sigma_{\text{int}}^{\alpha_4}$ (fb)   
& $\sigma_{\text{ano}}^{\alpha_4\alpha_4}$ (fb)
& $\sigma_{\text{int}}^{\alpha_5}$ (fb)
& $\sigma_{\text{ano}}^{\alpha_5\alpha_5}$ (fb)
& $\sigma_{\text{ano}}^{\alpha_4\alpha_5}$ (fb)
\\ 
\hline
$W^- [W^+ W^-] W^+$
& 2
& 4.27
& -45.9
& 6450.
& -25.4
& 8920.
& 13400.
\\
$W^- [W^+ W^-] W^+$
& 3
& 11.9
& -354.
& 89000.
& -224.
& 124$\times 10^3$
& 186$\times 10^3$
\\
$W^- [Z Z] W^+$
& 2
& 0.787
& 8.81
& 129.
& 31.0
& 771.
& 522.
\\
$W^- [Z Z] W^+$
& 3
& 2.77
& 68.4
& 1530.
& 235.
& 8390.
& 5650. 
\\
$[W^\pm Z] Z W^\mp$
& 2
& 0.596
& 2.88
& 98.2
& 8.46
& 263.
& 117.
\\
$[W^\pm Z] Z W^\mp$
& 3
& 1.17
& 11.2
& 1100.
& 31.3
& 1630.
& -240.
\end{tabular}
\end{center}
\caption{Values for the standard model, pure anomalous and interference cross
sections after cuts, according to Eq.\ (\protect\ref{crosssection}), for
$\gamma \gamma \to W^-W^+W^-W^+$ and $W^-ZZW^+$ and several $e^+e^-$
center--of--mass energies.  We display between brackets the vector bosons
produced in the central region.}
\label{sig:aa}
\end{table}

%
\begin{table}[hbt]
\begin{center}
\begin{tabular}{cccccccc}
events
& $\sqrt{s_{ee}}$ TeV 
& $\sigma^{lh}_{\text{sm}}$ (fb) 
& $F (WWWW)$
& $F (W[ZZ]W)$
& $F ([WZ]ZW)$
& $F ([WW]ZZ)$
& $F(t \bar{t})$
\\ 
\hline
$WW$ + 2 jets
& 2
& 0.34
& 50.4\% 
& 0.6\%
& 2.1\%
& 0.9\%
& 46.0\%
\\
$WW$ + 2 jets
& 3
& 0.51
& 80.2\%
& 1.0\%
& 2.5\%
& 0.8\%
& 15.5\%
\\
$ZZ$ + 2 jets
& 2
& 0.028
& 7.1\%
& 75.0\%
& 10.7\%
& 0.1\%
& 7.1\%
\\
$ZZ$ + 2 jets
& 3
& 0.071
& 8.4\%
& 82.5\%
& 7.0\%
& 0.7\%
& 1.4\% 
\end{tabular}
\end{center}
\caption{Total light Higgs background cross section in the $\gamma\gamma$ mode
  after cuts and detection efficiencies and its composition. We display
  between brackets the vector bosons produced in the central region.}
\label{bkg:aa}
\end{table}

%

\newpage


\newpage

\newpage

\end{document}